\documentclass[twocolumn,epsfig,letter]{jpsj3}

\usepackage{amsmath}
\usepackage{graphicx}

\def\nn{\nonumber}

\title{Polarization Dependence of Optical Transitions in Graphene Nanoribbons}

\author{
Ken-ichi \textsc{Sasaki}$^{1}$,
Keiko \textsc{Kato}$^{1}$,
Yasuhiro \textsc{Tokura}$^{1}$,
Tetsuomi \textsc{Sogawa}$^{1}$,
and
Riichiro \textsc{Saito}$^{2}$
}

\inst{
$^{1}$ NTT Basic Research Laboratories, 
Nippon Telegraph and Telephone Corporation,
3-1 Morinosato Wakamiya, Atsugi, Kanagawa 243-0198, Japan
\\
$^{2}$ Department of Physics, Tohoku University,
Sendai 980-8578, Japan
}

\recdate{\today}

\abst{
 The universality of $k$-dependent electron-photon and
 electron-phonon matrix elements is discussed for 
 graphene nanoribbons and carbon nanotubes.
 An electron undergoes 
 a change in wavevector in the direction of broken translational symmetry,
 depending on the light polarization direction.
 We suggest that this phenomenon originates from a microscopic feature of chirality.
}

\kword{graphene nanoribbon, armchair edge, zigzag edge, carbon nanotube,
chirality, optical selection rule, electron-phonon interaction}

\begin{document}

\maketitle

\begin{figure}[htbp]
 \begin{center}
  \includegraphics[scale=0.5]{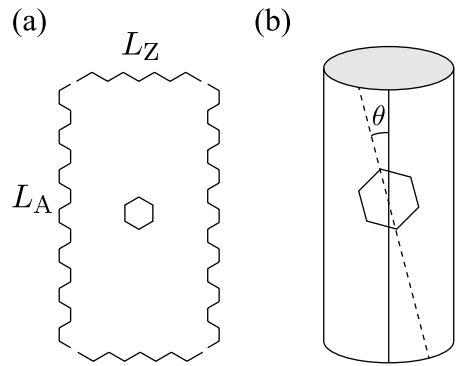}
 \end{center}
 \caption{(a) Chirality of a GNR. 
 A large aspect ratio $L_{\rm A}/L_{\rm Z} \gg 1$
 ($L_{\rm Z}/L_{\rm A} \gg 1$) defines an armchair (zigzag) GNR,
 where $L_{\rm A}$ and $L_{\rm Z}$ denote the lengths of the armchair
 and zigzag edge, respectively.
 (b) The chirality of a CNT is defined by the angle $\theta$ 
 with respect to the direction of the CNT axis.
 }
 \label{fig:topo}
\end{figure}

Recently, graphene nanoribbons (GNRs) have attracted attention 
for a variety of reasons.~\cite{areshkin07,han10,wang10}
GNRs are considered as unrolled carbon nanotubes (CNTs),
and a characteristic of GNRs (CNTs) 
is induced by the existence (absence) of edges.~\cite{tanaka87,klein94,fujita96}
GNRs as well as CNTs are categorized by chirality,~\cite{saito92apl}
and armchair GNRs (A-GNRs) and zigzag GNRs (Z-GNRs)
are known to have a high symmetry
[see Fig.~\ref{fig:topo}(a) and (b)].
Several experimental groups 
have attempted to clarify the optical properties of GNRs.~\cite{canifmmode04,xie11} 
Knowing the optical selection rule for GNRs will be 
an important step in understanding GNRs.
Theories suggest that A- and Z-GNRs exhibit 
different polarization dependence as regards 
optical absorption.~\cite{hsu07,gundra11,sasaki11-dc}  
Namely, 
the selection rule for Z-GNRs possesses a 90$^\circ$ rotation of
polarization with respect to the selection rule for A-GNRs.
This polarization dependence originates from the chirality dependent
electronic wavefunction for GNRs.~\cite{tanaka87,klein94,fujita96}
However, the derivations of the chirality dependence
described so far are complicated, 
inaccessible, 
and do not make it easy to grasp the essential point of the problem.
The purpose of this paper is to explain the selection rules
for A-GNRs, Z-GNRs, and CNTs in a unified manner. 
The procedure adopted in this paper is not only simple but also 
applicable to obtaining selection rules even 
for electron-phonon interactions.


First, we explain that the optical selection rule for A-GNRs 
is the same as that for CNTs. 
The electron-photon interaction is given by
${\bf A}({\bf r},t) \cdot \hat{\bf J}$, 
where ${\bf A}({\bf r},t)$ and $\hat{\bf J}$ 
are a vector potential and a current operator, respectively.
In the dipole approximation,
we assume a current operator of the form
$\hat{\bf J}=-(e/m)\hat{\bf p}$, 
where $-e$ is the electron charge, 
$m$ is the electron mass, and 
$\hat{\bf p}=-i\hbar \nabla$ is the momentum operator.
This current operator does not take the degree of sublattice (chirality)
into account.
Later we provide a proper matrix element for the current operator
that takes account of the chirality of graphene.
The importance of chirality will be clearly understood
with this line of argument.

Let us review CNTs.
As a result of the cylindrical geometry of CNTs,
the wavevector $(k)$ around the axis of a tube is a good quantum number,
and the wavefunction of an electron can be taken as a plane wave
if we omit the sublattice of graphene,
\begin{align}
 \psi_k(x) = \frac{1}{\sqrt{L}}e^{ikx},
 \label{eq:wf}
\end{align}
where $x$ and $L$ are the circumferential coordinate and length, respectively.
The wavevector is quantized by the periodic boundary condition 
$\psi_k(x+L)= \psi_k(x)$ as $k_n L = 2\pi n$ ($n$ are integers),
so that the electronic modes are labeled by the band index $n$ as $\psi_n(x)$.
The optical selection rule for the $n$th and $m$th states
is related to the possible change in wavenumber, $\Delta n\equiv m-n$.
The cylindrical geometry of CNTs
gives rise to a positional $x$-dependence for 
the vector potential $A_x$,~\cite{ajiki94} that is, 
for a light whose polarization direction is
perpendicular to the CNT axis,
\begin{align}
 A_x(x) = A \sin\left( \frac{2\pi}{L}x \right).
 \label{eq:standA}
\end{align}
Then we see that $\Delta n$ must be $\pm 1$ in order to have a non-zero
matrix element of $-(e/m)\oint A_x(x) \psi_m^*(x) \hat{p}_x
\psi_n(x)dx$.
This condition for wavenumber $\Delta n=\pm 1$ 
is the selection rule of CNTs 
with a perpendicularly polarized light.~\cite{ajiki94,ichida04,duesberg00}

With GNRs, 
a plane wave is reflected at the edge.
Consequently a standing wave is formed 
by the superposition of two plane waves
propagating in opposite directions.
Suppose that an incident wave has the wavevector $k$
in the direction perpendicular to the edge, 
then the wavevector of the reflected wave is given by $-k$ 
as a result of momentum conservation.
There are two possible types of superpositions:
$\psi_{k}(x)\pm\psi_{-k}(x)$.
For A-GNRs,
the electron wavefunction should vanish at the edge ($x=0$),
and only the antisymmetric combination,
\begin{align}
 \varphi_{k}(x) = \sqrt{\frac{2}{L}}\sin(k x),
 \label{eq:standwf}
\end{align}
is selected.
The boundary condition 
should be also imposed at $x=L$ as
$\varphi_{k}(L)=0$, which gives $k_{n} = \pi n/L$
($n$ are positive integers).
The standing waves are labeled by 
the band index $n$ as $\varphi_n(x)$.
The flat geometry of GNRs
results in a constant vector potential $A_x(x)=A$,
and the matrix elements are given by
$-(e/m)A \int_0^L \varphi_m(x) \hat{p}_x \varphi_n(x)dx$.
This integral results in
\begin{align}
  \frac{2}{L} & \int_0^L \sin(k_{m} x) \cos(k_{n} x) dx \nn \\
 &=
 \begin{cases}
  \displaystyle 0 & \text{$m-n \in$ even} \\
  \displaystyle 
  \frac{2}{\pi} \left( \frac{1}{m-n}+ \frac{1}{m+n} \right)
  & \text{$m- n \in$ odd}.
 \end{cases}
 \label{eq:int}
\end{align}
Because the Fermi wavevector satisfies $k_{\rm F} = 4\pi/3a$
($a$ denotes a lattice constant), 
the values of $m$ and $n$ are selected so that they are large enough for
$(m+n)^{-1}$ to be negligible 
compared with $(m-n)^{-1}$.
Thus, we approximate
$\frac{2}{L} \int_0^L \sin(k_{m} x) \cos(k_{n} x) dx = (2/\pi) \Delta
n^{-1}$.
Furthermore, the transition amplitudes for 
$\Delta n=\pm 3,\pm 5,\ldots$ are suppressed by the factor of $\Delta n^{-1}$.
The selection rule of A-GNRs is mainly given by $\Delta n =\pm1$,
which is coincident with the selection rule of CNTs
as discussed above.

It is meaningful to consider the reason for the similarity.
For CNTs, the change in electron wavenumber ($\Delta n =\pm1$)
is brought about by the macroscopic topology of the cylinder.
A light behaves as a standing wave on a cylindrical surface [eq.~(\ref{eq:standA})], 
while the electrons are plane waves [eq.~(\ref{eq:wf})].
For GNRs, the electrons are standing waves
[eq.~(\ref{eq:standwf})], while the light
is a uniform plane wave (i.e., zero-wavenumber mode).
An essential factor in obtaining $\Delta n =\pm1$ for A-GNRs
is that the momentum operator $\hat{p}_x$ alters
$\sin(kx)$ into $\cos(kx)$.
Since the momentum operator originates from 
a lattice spacing as 
$\varphi_n(x+a/2)-\varphi_n(x-a/2)\approx a (\partial/\partial
x)\varphi_n(x)$, we see that 
a microscopic lattice topology is essential for a change in 
the electron wavenumber.
Contrastingly, a plane wave 
satisfies $a(\partial/\partial x)\psi_n(x)\propto \psi_n(x)$,
and the orthogonality condition tells us that 
the corresponding current cannot change the electron wavenumber 
when $A_x(x)$ is constant.
Therefore, as long as standing waves are formed in GNRs,
we expect the wavnumber to change in an optical transition.
A change in the wavenumber will be a universal phenomenon 
for GNRs, as we see in the following.

\begin{figure}[htbp]
 \begin{center}
  \includegraphics[scale=0.3]{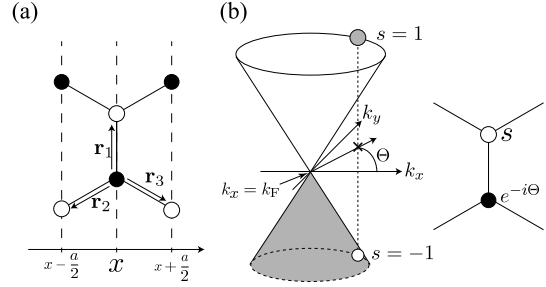}
 \end{center}
 \caption{(a) The bond vectors, ${\bf r}_1=a_{\rm cc}{\bf e}_y$, 
 ${\bf r}_2=a_{\rm cc}(-\frac{\sqrt{3}}{2}{\bf e}_x-\frac{1}{2}{\bf e}_y)$, and
 ${\bf r}_3=a_{\rm cc}(\frac{\sqrt{3}}{2}{\bf e}_x-\frac{1}{2}{\bf
 e}_y)$, where $a_{\rm cc}$ denotes the bond length.
 (b) The band index $s$ in the Dirac cone, and the relative amplitude
 between two sublattices is determined by $s$ and $\Theta$, respectively.
 These variables correspond to the amplitudes at two sublattices.
 }
 \label{fig:honeycomb}
\end{figure}


To investigate the polarization dependence of the selection rule,
we now take account of
the local arrangement of the carbon-carbon bonds
on an atomic scale [see Fig.~\ref{fig:honeycomb}(a)].
Since graphene's hexagonal unit cell
consists of two atoms (A and B),
the wavefunction has two components 
\begin{align}
 \Psi_{k_nk_ys}({\bf r}_{\rm A}) 
 &= 
\begin{pmatrix}
 \Psi^{\rm A}_{k_nk_y s}({\bf r}_{\rm A}) \cr 
 \Psi^{\rm B}_{k_nk_y s}({\bf r}_{\rm A}+{\bf r}_1)
\end{pmatrix}
 \nn \\
 &= \varphi_n(x) \psi_{k_y}(y)
 \frac{1}{\sqrt{2}}
 \begin{pmatrix}
  e^{-i\Theta(k_n,k_y)} \cr s
 \end{pmatrix},
\label{eq:wfa}
\end{align}
where $s$ is the band index ($+1$ for the conduction band, 
$-1$ for the valence band), ${\bf r}_{\rm A}=(x,y)$ denotes the position of an A-atom, 
${\bf r}_a$ ($a=1,2,3$) represents the bond vectors [see Fig.~\ref{fig:honeycomb}(a)],
and $\Theta$ is the polar angle in $(k_x,k_y)$ plane defined with
respect to the Dirac point at $(k_{\rm F},0)$, as shown in Fig.~\ref{fig:honeycomb}(b).
The current also consists of two components:
${\bf J}^{\rm A}({\bf r}_{\rm A})$ and ${\bf J}^{\rm B}({\bf r}_{\rm B})$, where
${\bf J}^{\rm A}({\bf r}_{\rm A})$ [${\bf J}^{\rm B}({\bf r}_{\rm B})$]
represents the electron's flow into an A-atom [B-atom] at ${\bf r}_{\rm
A}$ [${\bf r}_{\rm B}$]
from the nearest-neighbor B-atoms [A-atoms].
These components contribute to a local dipole moment and 
are essential to the optical transition.~\cite{grueneis03}
An optical transition in each sub-atom does not induce a local dipole
moment and can be omitted from the analysis.

Suppose that an incident light transfers 
an electron from the B-sites denoted by open circles 
in Fig.~\ref{fig:honeycomb}(a),
to the central A-site (solid circle).
The corresponding current component is given by 
\begin{align}
 {\bf J}^{\rm A}_{k_mk'_ys',k_nk_ys}({\bf r}_{\rm A}) 
 &= ie \frac{\gamma}{\hbar}
 \Psi^{\rm A}_{k_mk'_ys'}({\bf r}_{\rm A})^\dagger \times \nn \\
 &\left\{
 \sum_{a=1,2,3}
{\bf r}_a \Psi^{\rm B}_{k_nk_ys}({\bf r}_{\rm A}+{\bf r}_a)
 \right\},
 \label{eq:ja}
\end{align}
where $\gamma$ denotes the nearest-neighbor hopping integral.
By putting eq.~(\ref{eq:wfa}) into eq.~(\ref{eq:ja}),
we obtain 
$\int {\bf J}^{\rm A}_{k_mk'_y s',k_nk_ys}({\bf r})dy = \delta_{k'_yk_y}
\frac{s}{2} e^{i\Theta(k_m,k_y)}
{\bf j}^{\rm A}_{mn}(x)$, where 
\begin{align}
 {\bf j}^{\rm A}_{mn}(x)= & ie \frac{\gamma}{\hbar}
 \varphi_{m}(x) \times \Big\{ 
 {\bf r}_1 \varphi_{n}(x) + {\bf r}_2 e^{-ik_y (3a_{\rm cc}/2)}\varphi_{n}(x-\frac{a}{2}) \nn \\
 & 
 + {\bf r}_3 e^{-ik_y (3a_{\rm cc}/2)}\varphi_{n}(x+\frac{a}{2})
 \Big\}.
\label{eq:jsmal}
\end{align}
Because $e^{-ik_y (3a_{\rm cc}/2)}$ is approximately equal to $1$
for the states near the Dirac point, 
we set $e^{-ik_y (3a_{\rm cc}/2)}=1$ in eq.~(\ref{eq:jsmal}).~\cite{sasaki11-dc}
Thus, $k_y$ does not appear explicitly in the following analysis
for A-GNR, but it is implicitly taken into account through $\Theta(k_n,k_y)$.
Here let us define the amplitude for an A-atom of GNR
$[\alpha_i]_{mn} = \int A_i(x) {\bf e}_i \cdot {\bf j}^{\rm A}_{mn}(x)
dx$ ($i=x,y$), which are given by
\begin{align}
\begin{split}
 & [\alpha_x]_{mn} =iA ev_{\rm F}
 \frac{2}{L}\int_0^L \sin(k_m x) \cos(k_n x) dx, \\
 & [\alpha_y]_{mn} =iA ev_{\rm F} 
 \frac{2}{L}\int_0^L \sin(k_m x) \sin(k_n x) dx,
\end{split} 
\label{eq:alphaArm}
\end{align}
where $v_{\rm F}\equiv 3\gamma a_{\rm cc}/2\hbar$.
It is straightforward to show
$[\alpha_x]_{mn} = iev_{\rm F}A(2/\pi) \Delta n^{-1}$ 
by using eq.~(\ref{eq:int}),
and $[\alpha_y]_{mn} = iev_{\rm F}A\delta_{mn}$
owing to the orthogonality condition.
Similarly, 
the local electronic current component at a B-atom 
is written as
\begin{align}
 {\bf J}^{\rm B}_{k_mk'_ys',k_nk_ys}({\bf r}_{\rm B}) &= -ie \frac{\gamma}{\hbar}
 \Psi^{\rm B}_{k_mk'_ys'}({\bf r}_{\rm B})^\dagger \times \nn \\
 & \left\{\sum_{a=1,2,3}
 {\bf r}_a \Psi^{\rm A}_{k_nk_ys}({\bf r}_{\rm B}-{\bf r}_a)
 \right\},
 \label{eq:jb} 
\end{align}
where ${\bf r}_{\rm B}$ denotes the position of a B-atom.
We define the current and amplitude for a B-atom as
$\int {\bf J}^{\rm B}_{k_mk'_ys',k_nk_ys}({\bf r}_{\rm B})dy  = 
\delta_{k'_yk_y}
\frac{s'}{2} e^{-i\Theta(k_n,k_y)}
{\bf j}^{\rm B}_{mn}(x)$ and 
$[\beta_i]_{mn} = \int A_i(x) {\bf e}_i\cdot {\bf j}^{\rm B}_{mn}(x)dx$, respectively.
We obtain
\begin{align}
\begin{split}
 & [\beta_x]_{mn} =iA ev_{\rm F}
 \frac{2}{L}\int_0^L \sin(k_m x) \cos(k_n x) dx, \\
 & [\beta_y]_{mn} =-iA ev_{\rm F} 
 \frac{2}{L}\int_0^L \sin(k_m x) \sin(k_n x) dx.
\end{split} 
\label{eq:betaArm}
\end{align}

The optical transition amplitude 
for A-GNR is constructed from the sum of the current components, 
$\iint {\bf J}^{\rm A}_{k_mk_ys',k_nk_ys}({\bf r}_{\rm A})dxdy$
and 
$\iint {\bf J}^{\rm B}_{k_mk_ys',k_nk_ys}({\bf r}_{\rm B})dxdy$, as
\begin{align}
 & M_{i}(k_mk_ys',k_nk_ys) = \nn \\
 & \frac{1}{2} 
 \left( s e^{i\Theta(k_m,k_y)} [\alpha_i]_{mn}  
 + s' e^{-i\Theta(k_n,k_y)} [\beta_i]_{mn}  \right).
 \label{eq:matop}
\end{align}
By comparing eq.~(\ref{eq:alphaArm}) with eq.~(\ref{eq:betaArm}), 
we see that $[\beta]$ and $[\alpha]$ are related to each other via
\begin{align}
 [\beta_x]_{mn}=[\alpha_x]_{mn}, \ \ [\beta_y]_{mn}=-[\alpha_y]_{mn}.
\end{align}
Then, the matrix element of eq.~(\ref{eq:matop})
may be written in a more compact form (in $ev_{\rm F}A$ units) as
\begin{align}
\begin{split}
 & M_{x}(k_mk_ys',k_nk_ys)=
 \frac{i}{\pi \Delta n} 
 \left( s e^{i\Theta(k_m,k_y)}  
 + s' e^{-i\Theta(k_n,k_y)} \right), \\
 & M_{y}(k_mk_ys',k_nk_ys)
 = \delta_{mn}\frac{i}{2} \left( s e^{i\Theta(k_m,k_y)} - s' e^{-i\Theta(k_n,k_y)} \right).
\end{split}
\label{eq:amat}
\end{align}
Equation~(\ref{eq:amat}) is the optical matrix elements for A-GNRs.
This result was used to explain the optical absorption spectra
in ref.~\citen{sasaki11-dc}.

The matrix elements for Z-GNRs are obtained by repeating a similar
calculation to that given above.
A standing wave for Z-GNRs is formed 
by the superposition of two waves
propagating in opposite $y$-directions.
Suppose that an incident wave has a wavevector $k_y$
in a direction perpendicular to the edge, 
then the wavevector of the reflected wave is given by $-k_y$ 
as a result of momentum conservation.
The possible superpositions are 
\begin{align}
 \psi_{k_y}(y)
 \frac{1}{\sqrt{2}}
 \begin{pmatrix}
  e^{-i\Theta(k_x,k_y)} \cr s
 \end{pmatrix}
 \pm 
 \psi_{-k_y}(y)
 \frac{1}{\sqrt{2}}
 \begin{pmatrix}
  e^{-i\Theta(k_x,-k_y)} \cr s
 \end{pmatrix}.
\end{align}
Here, we assume that the zigzag edges appear at $y=0$ and $y=L$, and
consider a case where the zigzag edge at $y=0$ ($y=L$)
consists of A-atoms (B-atoms).
Then, the electron wavefunction at the B-atoms
should vanish at the edge ($y=0$),
and the antisymmetric combination is selected.
Thus, the standing wave for Z-GNRs is written as
\begin{align}
 \Psi_{k_xk_ys}(x,y) = \psi_{k_x}(x) \frac{1}{\sqrt{L}}
 \begin{pmatrix}
  \sin[k_y y-\Theta(k_x,k_y)] \cr s \sin(k_y y)
 \end{pmatrix}.
\label{eq:wfz}
\end{align}
The boundary condition 
should also be imposed at $y=L$ as
$\Psi^{\rm A}_{k_xk_ys}(x,L)=0$, 
by which $k_y$ is quantized according to $k_{n}L-\Theta(k_x,k_{n})=n\pi$.
Using the wavefunction $\Psi_{k_xk_ns}(x,y)$, 
we obtain the optical matrix element,~\cite{sasaki11-dc}
\begin{align}
\begin{split}
 & M_{x}(k_xk_ms',k_xk_ns) \\
 = &\frac{i}{2} 
 \left( s \cos\Theta_m + s' \cos\Theta_n \right) [\alpha_y]_{mn} \\
 &-\frac{i}{2} 
 \left( s \sin\Theta_m [\alpha_x]_{nm} + s' \sin\Theta_n [\alpha_x]_{mn}
 \right), \\
 & M_{y}(k_xk_ms',k_xk_ns) \\
 =&\frac{1}{2} 
 \left( s \cos\Theta_m - s' \cos\Theta_n \right) [\alpha_y]_{mn} \\
 &-\frac{1}{2} 
 \left( s \sin\Theta_m [\alpha_x]_{nm} - s' \sin\Theta_n [\alpha_x]_{mn}
 \right),
\end{split}
\label{eq:zmat}
\end{align}
where we abbreviate $\Theta(k_x,k_n)$ as $\Theta_n$ and 
use the fact that $k_x \approx k_{\rm F}$ to obtain the right-hand side.
For an inter-band transition ($s=-1$ and $s'=1$),
we see that the wavenumber selection rule 
is given by $\Delta n= \pm1,\pm3,\ldots$ ($\Delta n=0$)
when the polarization of an incident light 
is set parallel (perpendicular) to the zigzag edge, 
which is a 90$^\circ$ rotation of polarization
with respect to the selection rule for A-GNRs.
For an intra-band transition ($s=s'$),
the selection rule for wavenumber 
exhibits a 90$^\circ$ rotation of polarization
with respect to the $\Delta n$-selection rule 
for an inter-band transition.
This additional change in the polarization dependence
is a characteristic feature specific to Z-GNRs, 
not shown in the $\Delta n$-selection rule for A-GNRs.

In Table~\ref{tab:sum},
we show our results for the electron-photon matrix elements for GNRs,
where we pay special attention to the change in wavenumber $\Delta n$.
By replacing $\varphi_n(x)$ in eq.~(\ref{eq:wfa})
with $\psi_n(x)$, we see that the polarization dependence of the
$\Delta n$-selection rule for CNTs 
is irrelevant to the chirality.
The existence of the edge and 
the resultant formation of the chirality-dependent standing wave 
are important to obtaining the chirality-dependent
$\Delta n$-selection rule.

\begin{table}[htbp]
 \caption{
 Polarization dependence of $\Delta n$ for the possible inter-band optical transition. ``para'' (``perp'')
 means that the polarization is parallel (perpendicular)
 to the edge (axis) of a GNR (CNT). 
 For GNRs, $\Delta n$ induced by an electron-phonon interaction
 follows the same rule, where ``para'' (``perp'')
 means the direction of the vibration.
 \label{tab:sum}
 }
 \begin{tabular}{c|ccc}
  \hline 
   & A-GNRs & Z-GNRs & CNTs \\
   \hline   
   & \\
   para & $\Delta n=0$ & $\Delta n = \pm1,\pm3,\ldots$ & $\Delta
	       n=0$ \\
   & \\
   \hline 
   & \\
   perp & $\Delta n=\pm1,\pm3,\ldots$ & $\Delta n=0$ & $\Delta n=\pm1$\\
   & \\
  \hline
 \end{tabular}
\end{table}

Finally, we briefly mention that 
this method is applicable to an electron-phonon interaction.
A relative displacement vector for an optical phonon mode ${\bf u}({\bf r},t)$
($\equiv {\bf u}_{\rm A}-{\bf u}_{\rm B}$) 
changes the hopping integral from $\gamma$ 
to $\gamma+g{\bf u}({\bf r},t) \cdot ({\bf r}_a/a_{\rm cc})$, 
where $g$ is a coupling constant, and 
induces a deformation current.
The current induced at an A-atom by the lattice deformation ${\bf u}({\bf r})$
is given as ${\bf J}^{\rm A}_{{\bf k}'s',{\bf k}s}({\bf r}_{\rm A})$ 
by replacing $-e\gamma$ with $g/a_{\rm cc}$.
The current induced at a B-atom is given by 
$-{\bf J}^{\rm B}_{{\bf k}'s',{\bf k}s}({\bf r}_{\rm B})$, where 
the extra minus sign originates from the fact that 
${\bf u}({\bf r},t)$ changes its sign when we replace A and B-atoms:
the interaction must be symmetric with respect to the change of A and B
as $({\bf J}^{\rm A}-{\bf J}^{\rm B})\cdot ({\bf u}_{\rm A}-{\bf u}_{\rm
B})$.
Therefore, the transition amplitude is constructed 
from the difference between the current components,
\begin{align}
 \iint \left\{  {\bf J}^{\rm A}_{{\bf k}'s',{\bf k}s}({\bf r}_{\rm A}) - 
 {\bf J}^{\rm B}_{{\bf k}'s',{\bf k}s}({\bf r}_{\rm B}) \right\}dxdy.
\end{align}
Since the minus sign is taken into account as
$s' \to -s'$ in Eqs.~(\ref{eq:amat}) and (\ref{eq:zmat}),
we conclude that the electron-phonon matrix element for 
an intra-band (inter-band) transition is the same as the electron-photon matrix
element for an inter-band (intra-band) transition, 
except for a numerical factor.
Thus, 
the selection rule for the wavenumber of a phonon mode 
whose vibrational direction is either parallel or perpendicular 
to the edge
allows a non-zero shift $\Delta n={\rm odd}$
in the wavenumber of an electron (see Table~\ref{tab:sum}).

In conclusion, 
the electronic standing waves in GNRs undergo a change in wavenumber
through an optical, inter-band transition when the polarization of an incident light
is perpendicular (parallel) to the armchair (zigzag) edge.
The origin of the shift in wavenumber is attributed to the 
microscopic topology of the lattice, in that
the standing wave has different probability amplitudes at
adjacent sites in a chirality dependent manner.
Thus, a change in wavenumber is a universal phenomenon in GNRs.
The universality is also recognized for electron-phonon interactions.

K.S acknowledges a MEXT Grant (No.~23310083).
R. S. acknowledges a MEXT Grant (No.~20241023).

\bibliographystyle{./jpsj}
%

\end{document}